\documentstyle[12pt,fleqn,cite,epsfig]{article}

\def\be{\begin{equation}}
\def\ee{\end{equation}}
\def\bea{\begin{eqnarray}}
\def\eea{\end{eqnarray}}

\begin{document}
\begin{titlepage}
\begin{center}
\hfill hep-th/0002150\\
\hfill CERN-TH/2000-059\\
\vskip .2in

{\Large \bf Non-Planar String Networks on Tori}
\vskip .5in

{\bf Alok Kumar$^{1,2}$\\
\vskip .1in
{\em 1. Theory Division, CERN,\\
CH-1211, Geneva 23, Switzerland}\\
{\em 2. Institute of Physics,\\
Bhubaneswar 751 005, INDIA\\email: Alok.Kumar@cern.ch}}
\end{center}

\begin{center} {\bf ABSTRACT}
\end{center}
\begin{quotation}\noindent
\baselineskip 10pt
Type II strings in $D=5$ contain particle-like $1/8$ supersymmetric
BPS states. In this note we give a string-network representation of 
such states by considering (periodic) non-planar $(p,q,r)$-string networks 
of eight dimensional type II string theory on $T^3$. 
We obtain the BPS mass formula of such states, in terms of charges  
and generating-vectors of the torus, and 
show its invariance under an $SL(3, Z)\times SL(3, Z)$ group of 
transformations. Results are then generalized to string-networks
associated with the $SL(5, Z)$ $U$-duality in seven dimensions.
We also discuss reinterpretation
of the above $(D=5)$ mass formula in terms of BPS states in 
world-volume theories of $U2$-branes in $D=8$. 
\end{quotation}
\vskip .2in
CERN-TH/2000-059\\
February 2000\\
\end{titlepage}
\vfill
\eject


\section{Introduction}
Classification of type II BPS states with different supersymmetries
have been discussed in several papers\cite{hull,ferrara}. 
They form $U$-duality multiplets of 
stings in various dimensions. The
corresponding black hole entropies are also 
given by duality invariant expressions. However, 
its statistical understanding requires the 
degeneracy of states. It has been suggested\cite{sen}
that string networks\cite{sen,bhatt,kuma} 
may provide the degeneracy of such states, 
when considered on tori\cite{sen}.  

Network configurations have also been an important topic of 
discussion from various other angles in type II string 
theory\cite{schwarz,dasg,zwieb}, 
$N=4$ gauge theories\cite{bergman,bergman2,4d} and non-commutative
geometry\cite{non-com}. In string theory,
they provide $1/4$, $1/8$ and other lower supersymmetric 
states\cite{schwarz,dasg,zwieb,sen,bhatt,kuma}.
Moreover type IIB planar string networks 
which end on D3-branes\cite{bergman,bergman2} 
represent the $1/4$ BPS dyonic states of 
$N=4$ gauge theories in four dimensions\cite{4d}. 
Network-like configurations,
have also appeared in other supersymmetric field theories\cite{saffin}, 
but are likely to have connections with those mentioned above.
Recently, extension of network configurations to strings
carrying 1-form electric charges (per unit length) and currents 
was also presented\cite{kuma}. These can be of interest from the point of 
view of cosmological applications\cite{sudip}.

In this paper we study the
application of string networks to $D=5$ BPS states.
It is known that one can have $1/8$ supersymmetric 
particle-like states in $D=5$\cite{ferrara}. 
We give a string network representation of 
such states, by compactifying periodic
$D=8$ non-planar networks\cite{bhatt} of an  
$SL(3,Z)$-multiplet of type II strings
on $T^3$. We also generalize the results to the $SL(5,Z)$ 
U-duality in seven dimensions.

Our exercise can also be used to write down 
mass formula of $1/8$ BPS states, in certain 
world-volume theories of 2-branes. These branes are themselves 
identified as $U$-duality branes\cite{liu} 
obtained from toroidal 
compactification of ten-dimensional branes. For the 
eight dimensional case, one notices 
that there exist 2-branes
which are invariant under the $SL(3, Z)$ part of $U$-duality 
(for the purpose of this paper, $SL(3, Z)$ is the only 
relevant part of $U$-duality in eight dimensions). Masses of 
states, which can be identified as a string-junction ending on 
these branes can then be found from the above exercise. 
Such an analysis for D3-branes has been 
performed in great detail, including for the case
of non-abelian gauge groups etc.\cite{bergman,bergman2,4d}. 
We expect that, same 
should be possible for these $U2$-branes as well. 

We now start by describing the periodic non-planar
$(p,q,r)$ string networks of our interest, built out
of basic structures as shown in figures-1. 
For convenience, we first consider 
the string networks, whose basic building blocks are 
4-string junctions as in figure-1(a).
The existence of such a junction can be seen from the general 
structure of non-planar string networks of \cite{bhatt}. In 
the construction of \cite{bhatt}, the basic building blocks of the 
networks are 3-string junctions whose 3-prongs lie in a specific two 
dimensional plane in a three dimensional space, now identified as 
$T^3$. However, different junctions, including the adjacent ones
can have their 3-prongs in different two dimensional planes, 
giving them a non-planar form (see figure-1(b)). Then
by shrinking the length of the intermediate 
links of such adjacent junctions, which is a free parameter in  
these BPS constructions, one gets a 4-string junction.  Such objects
have also been studied in \cite{roy}. 
\begin{figure}[htb]
\epsfxsize=4in
\centerline{\epsffile{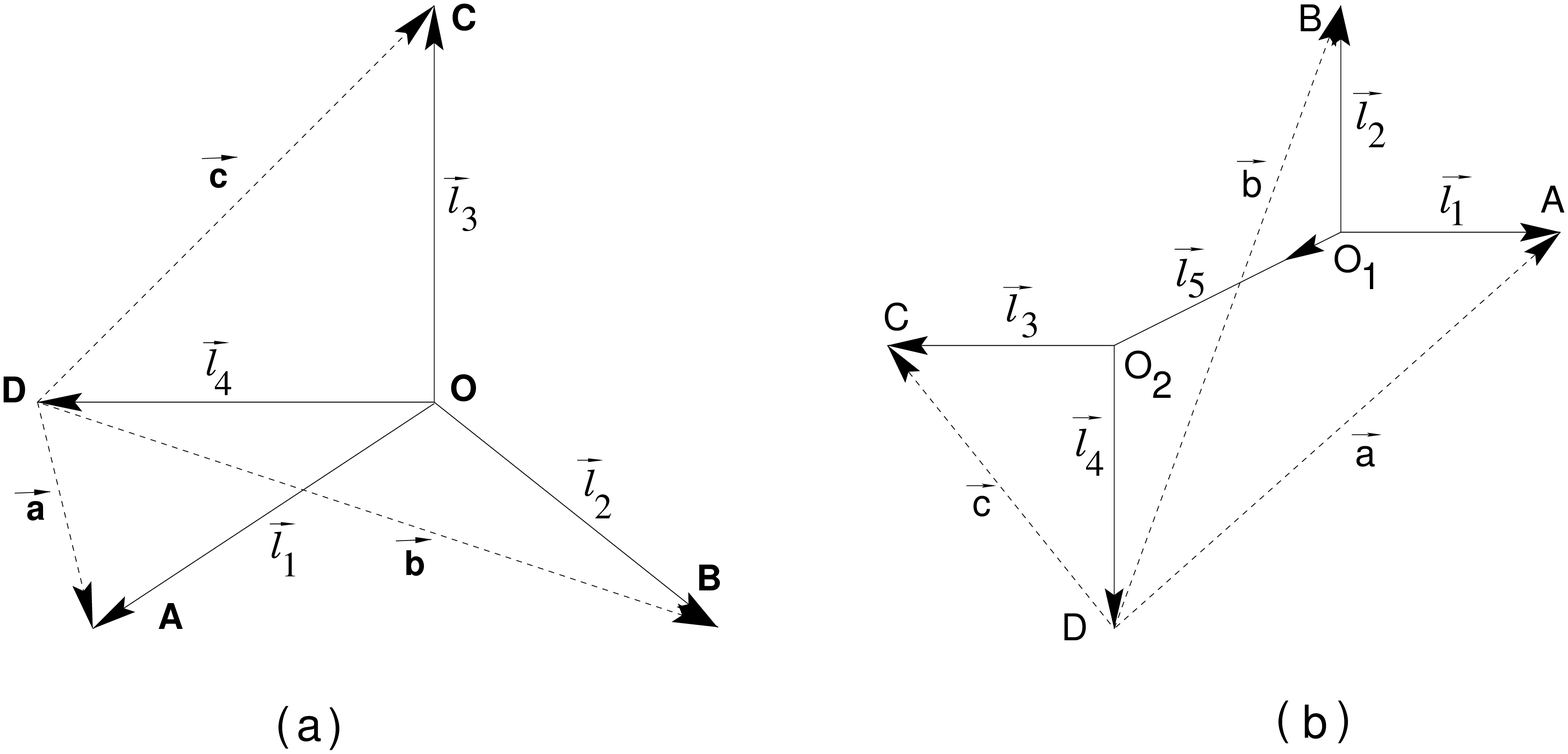}}
\caption{\baselineskip=11pt (a) {\em\small A 4-string junction.}~~~~ 
(b) {\em\small A non-planar 4-prong.}} 
\end{figure}
A periodic structure of such 4-string junctions can then be 
constructed as a three dimensional generalization of the string 
network lattice
in \cite{sen}. However one now needs
four strings with $SL(3)$ charges: $(p_I, q_I, r_I)$, $I=1,2,3,4$
to construct such 4-string junctions.
By fixing the lengths of these string-links to $l_I$, $(I=1,2,3,4)$,
and by imposing charge conservation condition on 
junctions: $\sum_{I=1}^4 p_I$ =  
$\sum_{I=1}^4 q_I$ =  $\sum_{I=1}^4 r_I$ = $0$, one obtains 
a periodic lattice. Although we
do not present a pictorial representation of such 3-D periodic
networks, their existence is guaranteed from the existence of 
the three dimensional lattice vectors $\vec{a}$, $\vec{b}$
and $\vec{c}$ given below in terms of the `link-vectors' $\vec{l}_I$.
These link-vectors themselves are given by the lengths of the 
prongs mentioned above and their 
orientation is given as in \cite{bhatt} in order to preserve 
$1/8$ supersymemtry. More precisely, these orientations for 
a string with $SL(3, Z)$ quantum numbers ${(P_I)}_i \equiv (p_I, q_I, r_I)$
are given in terms of components ${(X_I)}_a$, $(a=1,2,3)$ of a 
vector in real space (now identified as $T^3$): 
\begin{equation}
\vec{V_I} = {(X_I)}_a \hat{e}_a, \>\>(a=1,2,3), 
\label{defvec}
\end{equation}
with ${(X_I)}_a$ given by:
\begin{equation}
  {(X_I)}_a = (\lambda^{-1})_{a i} {(P_I)}_i. 
\label{defx}
\end{equation} 
`$\hat{e}_a$' in this paper always denotes orthogonal set of unit vectors in 
$T^3$, although its index $`a'$ is chosen to be same as that of an internal 
$SO(3)$ vector. $\lambda^{-1}$ in the above equation 
denotes the vielbein corresponding to the 
$SL(3)/SO(3)$ moduli:
\begin{eqnarray}
G = \pmatrix{g+ a^T a e^{-\phi} & e^{-\phi}a^T\cr
a e^{-\phi} & e^{-\phi}}, 
\label{modG}
\end{eqnarray}
with $g$ being a $2\times 2$ matrix:
\begin{eqnarray}
g = \pmatrix{e^{(\phi+\alpha)} + \chi^2 e^{-\alpha} & 
e^{-\alpha} \chi \cr \chi e^{-\alpha} & e^{-\alpha}}. 
\label{modgsmall}
\end{eqnarray}
Then one has 
\begin{eqnarray}
\lambda^{-1} = \pmatrix{e^{-(\phi+\alpha)/2} & 
-\chi e^{-(\phi+\alpha)/2} & - e^{-(\phi+\alpha)/2}a_1 + 
\chi e^{-(\phi+\alpha)/2}a_2 \cr
0 & e^{\alpha/2} & - e^{\alpha/2} a_2 \cr
0 & 0 & e^{\phi/2}}. 
\label{vielbein-1}
\end{eqnarray}
To summarize these definitions, $P_I$'s defined above are  
internal $SL(3, Z)$ vectors associated with the $I$'th string,
whereas $X_I$'s are internal $SO(3)$ vectors constructed by contracting 
$P_I$'s with the vielbein $\lambda$. This $SO(3)$ is the maximal 
compact subgroup of $SL(3)$. 
Finally $\vec{V}_I$'s in eqn.(\ref{defvec}) 
are vectors in $T^3$, due to their dependence on unit vector
$\hat{e}_a$. Identification of its components with those 
of $X_I$'s in eqns. (\ref{defvec}), (\ref{defx}) is a property 
of the string networks, as the spatial and internal 
orientations of the links in a network
are always aligned in a specific manner.

Major exercise now is to start from the expression of the
mass associated with the 
above 4-string junction defining the unit cell\cite{footnote} of a 
periodic network lattice and to show that these can be rewritten in terms
of three independent $SL(3, Z)$ charges $(P_I)_i \equiv (p_I, q_I, r_I)$, 
$(I=1, 2, 3)$, $SL(3)/SO(3)$ moduli $G$, and 
three-dimensional vectors: $\vec{a}, \vec{b}, \vec{c}$ 
defined in terms of the lengths $l_I$'s of the four legs of
the string junction as well as the unit vectors along these
legs, $\hat{n}_I \equiv {\vec{V}_I/|V_I|}$: 
\begin{equation}
 \vec{a} = \vec{l}_1 - \vec{l}_4,\>\>
\vec{b} = \vec{l}_2 - \vec{l}_4,\>\>
\vec{c} = \vec{l}_3 - \vec{l}_4. \label{vec-abc}
\end{equation}
In fact, as we will see below various combination of 
$(\vec{a},\vec{b}, \vec{c})$
provide additional moduli in the lower dimensional 
theory, after quantum numbers $(p_4, q_4, r_4)$ are eliminated 
in favor of the remaining ones, using charge conservation conditions. 

Technical non-triviality of our exercise, with respect to the one performed
in \cite{sen} for the planar network, is in dealing with 
the 3-dimensional problem in our case, compared to the 2-dimensional 
one in \cite{sen}. To perform this exercise explicitly, we first 
consider the case when $SL(3)/SO(3)$ moduli, $G$, have a diagonal form. 
It will be observed that the final expression  that we derive,
easily generalize to the most general moduli as well. For the 
diagonal case, $G$ has a form:
$G = diag.( e^{\phi + \alpha}, e^{-\alpha}, e^{-\phi})$.
Moreover for this case, the string tension is given as:
$T_I = |X_I| = [e^{-(\phi +\alpha)} p_I^2 + e^{\alpha} q_I^2 
         + e^{\phi} r_I^2 ]^{1\over 2}, \>\> (I=1,2,3,4)$.

We now use the above expressions to compute the mass of the BPS state,
given by the string network configuration built by 
the above 4-string junctions. It is given by 
\begin{equation}
m_{BPS}^2 = (l_1 T_1 + l_2 T_2 + l_3 T_3 + l_4 T_4)^2. 
                          \label{mass}
\end{equation}
Now,  to eliminate the lengths of the link-vectors 
of the strings in favor of the 
generating vectors $\vec{a}, \vec{b}, \vec{c}$, we use
expressions of various scalar and vector combinations formed 
from these by taking their dots and cross products:
$a^2 = l_1^2 + l_4^2 - 2 \vec{l}_1.\vec{l}_4,\>
\vec{a}.\vec{b} = \vec{l}_1.\vec{l}_2 - \vec{l}_1.\vec{l}_4 
                 - \vec{l}_2.\vec{l}_4 + l_4^2, \>
\vec{a}\times\vec{b} = \vec{l}_1\times \vec{l}_2 - 
\vec{l}_1\times \vec{l}_4 - \vec{l}_4 \times \vec{l}_2 $
etc.. Moreover, these expressions can be rewritten 
in terms of quantum numbers $(p_I, q_I, r_I)$, $(I=1,2,3)$, moduli
fields ($\phi$, $\alpha$), lengths  of the links
$l_I$, and their string-tensions $T_I$, by using relations:
\begin{eqnarray}
\vec{l}_1\times \vec{l}_2  = {l_1 l_2\over {T_1 T_2}}
\left[\hat{e}_1(q_1 r_2
- q_2 r_1) e^{{(\phi+\alpha)}/2}  +  \right. & \cr
\hat{e}_2(r_1 p_2
- r_2 p_1) e^{{-\alpha}/2} + 
\hat{e}_3(p_1 q_2
- p_2 q_1) e^{{-\phi}/2} 
\left. \right], &  
\label{l1l2cross}
\end{eqnarray}
and two other expressions obtained by taking cyclic permutations
in indices $(1, 2, 3)$ and $(p, q, r)$. Similarly,
\begin{eqnarray}
\vec{l}_1\times \vec{l}_4  = - {l_1 l_4\over {T_1 T_4}}\left[\hat{e}_1
(q_1 (r_2 + r_3) - (q_2 + q_3) r_1) e^{{(\phi+\alpha)}/2}  + \right. & \cr  
 \hat{e}_2(r_1 (p_2 + p_3) 
- (r_2+r_3) p_1) e^{{-\alpha}/2} + 
\hat{e}_3(p_1 (q_2 + q_3)
- (p_2 + p_3) q_1) e^{{-\phi}/2} 
\left. \right],
\label{l1crossl4}
\end{eqnarray}
and again two others obtained by the above cyclic permutations.

With the help of above expressions, and after some algebra, 
one can show that the mass of the BPS state, after $T^3$ compactification
can be written as:
\begin{eqnarray}
m_{BPS}^2 = \left[(\vec{V}_1.\vec{V}_1) a^2 + (\vec{V}_2.\vec{V}_2) b^2 + 
(\vec{V}_3.\vec{V}_3) c^2 + 2 (\vec{V}_1.\vec{V}_2)(\vec{a}.\vec{b})
+ 2 (\vec{V}_1.\vec{V}_3)(\vec{a}.\vec{c})
+ \right. & \cr 
2 (\vec{V}_2.\vec{V}_3)(\vec{b}.\vec{c}) 
\left. \right] 
+ 2 \left[(\vec{a}\times\vec{b}).(\vec{V}_1\times\vec{V}_2) +
(\vec{a}\times\vec{c}).(\vec{V}_1\times\vec{V}_3) +
(\vec{b}\times\vec{c}).(\vec{V}_2\times\vec{V}_3)\right], \cr
\equiv m_1^2 + m_2^2,  
\label{finalmass}
\end{eqnarray}
where $m_1^2$ and $m_2^2$ correspond to the terms in the two 
square brackets in eqn.(\ref{finalmass}). 
This equation is one of the main result of this paper. It gives the BPS mass
in terms of nine integers $(p_I, q_I, r_I)$'s, moduli $(\phi, \alpha)$
(through their appearance in $\vec{V_I}$), as well new 
set of moduli formed out of $\vec{a}, \vec{b}, \vec{c}$. 
The generalization of the result, to the case when the 
full set of $SL(3)/SO(3)$ 
moduli are turned on, is straight-forward. 
In that case, mass formula remains same as (\ref{finalmass}). 
However $\vec{V}_I$'s and $X_I$'s  involve general $SL(3)/SO(3)$
moduli through their dependence on the vielbein in 
eqn.(\ref{vielbein-1}).

We now show that the above mass formula has an 
$SL(3, Z)_U\times SL(3, Z)_u$
symmetry. The first $SL(3, Z)$ is essentially the $U$-duality 
symmetry of type II strings in eight dimensions. The second $SL(3, Z)$
comes from the compactification of the network on $T^3$. 

We first show the $SL(3, Z)\times SL(3, Z)$ invariance of the terms 
in the first square bracket in  eqn.(\ref{finalmass}), identified
as $m_1^2$. These terms can be rewritten as:
\begin{equation}
m_1^2 = P^T M P, 
\label{bpsmass1}
\end{equation}
where $P$ is $9\times 1$ column vector with entries:
\begin{eqnarray}
P = \pmatrix{P_1\cr P_2\cr P_3},
\end{eqnarray}
and $M$ is a matrix:
\begin{eqnarray}
M = \pmatrix{ a^2{G^{-1}} & ({\vec{a}.\vec{b}}){G^{-1}}  
& ({\vec{a}.\vec{c}}){G^{-1}} \cr
({\vec{b}.\vec{a}}){G^{-1}}  & b^2{G^{-1}} &  
({\vec{b}.\vec{c}}){G^{-1}} \cr 
({\vec{c}.\vec{a}}){G^{-1}} & ({\vec{c}.\vec{b}}){G^{-1}} 
& c^2{G^{-1}}}. 
\label{bigmoduli}
\end{eqnarray}

One can then write down the action of two $SL(3)$'s mentioned above
on charges and moduli, including the ones constructed out of vectors
$\vec{a}, \vec{b}, \vec{c}$. $SL(3)_U$ has the identical action as
in eight dimensions. This leaves any $T^3$ vectors such as 
$\vec{a}, \vec{b}, \vec{c}$ etc. invariant and acts on $P$ through 
a diagonal action on $P_I$'s as:
\begin{equation}
P_I \rightarrow \Lambda_U P_I,\>\>\>
G^{-1} \rightarrow {\Lambda_U^{-1}}^T G^{-1} \Lambda_U^{-1}.
\label{sl3U}
\end{equation}
with $\Lambda_U$ being $SL(3, Z)$ matrices. 
Second symmetry, namely $SL(3)_u$ acts on $P$ as:
\begin{eqnarray}
\pmatrix{P_1 \cr P_2 \cr P_3} \rightarrow 
\Lambda_u \pmatrix{P_1 \cr P_2 \cr P_3}, 
\label{vecP}
\end{eqnarray} 
namely, it mixes the indices $(1, 2, 3)$ associated with $SL(3)$ charges
$(p, q, r)$ of various strings 
among themselves. In addition, one has
\begin{eqnarray}
\vec{A} \equiv \pmatrix{\vec{a}\cr \vec{b}\cr\vec{c}} \rightarrow 
{\Lambda_u^{-1}}^T\pmatrix{\vec{a}\cr \vec{b}\cr\vec{c}}. 
\label{vecA}
\end{eqnarray}
Due to the action of the symmetry group defined above, 
$9\times 9$ moduli matrix in the compactified 
theory $(M)$, transforms under $SL(3)$'s as
$M \rightarrow (\Lambda_U^{-1 T}\otimes I_3) M 
(\Lambda_U^{-1}\otimes I_3), \>
M \rightarrow (I_3\otimes \Lambda_u^{-1 T}) M 
(I_3\otimes \Lambda_u^{-1})$.

We have therfore shown an explicit $SL(3, Z)_U\times SL(3, Z)_u$ invariance of
the first part of the BPS mass formula (\ref{bpsmass1}). We also observe
that by factoring out the volume of the polyhedron formed out of 
vectors $(\vec{a}, \vec{b}, \vec{c})$ from the matrix $M$ in 
(\ref{bigmoduli}), it can be identified with a matrix 
parameterizing $[SL(3/SO(3)] \times [SL(3)/SO(3)]$ moduli.  

We now see that 
the terms in the second square bracket of (\ref{finalmass}), namely
$m_2^2$, are also
invariant under the transformations (\ref{sl3U}), (\ref{vecP})
and (\ref{vecA}). We first consider
$SL(3)_U$, after writing vectors $\vec{V}_I$'s as:
$\vec{V}_I = T_I \hat{n}_I, \>\> (I = 1,2,3)$,
with $T_I$ being the tensions of the strings and $\hat{n}_I$ 
being the unit vectors along them. $SL(3, Z)_U$ invariance of 
$m_2^2$ then follows from the fact that it acts 
on various quantities inside second square bracket in 
(\ref{finalmass}) only
through terms in the expressions of string tensions.

$SL(3, Z)_u$ symmetry of $m_2^2$
is also clear by noticing that although $\vec{V}_I$'s are
spatial (or $T^3$) vectors, they transform 
under $SL(3, Z)_u$ due to its action on 
quantum numbers $p_I, q_I, r_I$'s in a similar manner as
$P_I$'s mentioned above in (\ref{vecP}). Then, using the 
definition $\vec{A}$ as in (\ref{vecA}),  
the invariance of $m_2^2$ can be seen by writing it as:
\begin{equation}
m_2^2 = (\vec{A}_I \times \vec{A}_J ). (\vec{V}_I \times \vec{V}_J). 
\label{bpsmass2}
\end{equation}
We have therefore shown an $SL(3, Z)\times SL(3, Z)$ invariance of the 
mass formula obtained from a periodic network of 4-string 
junctions in eight dimensions.

A similar analysis goes through for more general periodic string-networks
constructed out of 4-prong structures as shown in figure-1(b). 
One now has lattice vectors defined as:
\begin{equation}
 \vec{a} = \vec{l}_1 - (\vec{l}_4 + \vec{l}_5),\>\>
\vec{b} = \vec{l}_2 - (\vec{l}_4 + \vec{l}_5),\>\>
\vec{c} = \vec{l}_3 - \vec{l}_4. \label{vec-abc2}
\end{equation}
The mass of the $1/8$ supersymmetric BPS state associated with 
the compactified string network is now given by the expression:
\begin{equation}
m^2_{BPS} = (\sum_{i=1}^5 l_I T_I)^2, 
\label{newmass}
\end{equation}
with lengths and tensions now being associated 
with the string-links in figure-1(b).
This expression, after similar algebra as above, can now be written as:
\begin{equation}
m^2_{BPS} = P^T M P + \sum_{I=1}^5 (\vec{l}_I\times \vec{l}_J). 
                                   (\vec{V}_I\times \vec{V}_J) .
\label{prongmass}
\end{equation}
Then using the definitions of the lattice vectors in (\ref{vec-abc2}),
and charge-conservations on vertices $O_1$ and $O_2$ in 
figure-1(b), one can show that the final $1/8$ BPS mass formula
is once again given by eqns. (\ref{finalmass}), (\ref{bpsmass1})
and (\ref{bpsmass2}).

We also obtain the $SL(2, Z)\times SL(2, Z)$ invariant formula
of \cite{sen} by turning off appropriate moduli and charges. For 
example, when only nonzero $SL(3)$ charges are: $(p_I, q_I)$, 
$(I=1, 2)$, then by setting 
$\phi= a_1 = a_2 =0$ in (\ref{finalmass}) one reproduces 
exactly the same expression as in \cite{sen}. 
This can be seen from the form of $m_1^2$ in (\ref{bpsmass1}), 
which reduces to the first term in eqn.(17) of \cite{sen}
in these limits. Moreover, for $m_2^2$ only one of the 
term in the second square bracket in (\ref{finalmass}) is 
nonzero and gives precisely the second term in 
eqn.(17) of \cite{sen}. 

We now comment on the connection of these results with  
$U$-duality in $D=5$. The full $U$-duality symmetry 
in $D=5$ is $E_{6 (6)}$ and 
gauge charges are in its {\bf 27}-dimensional 
representation. $E_{6 (6)}$ however has an $SL(6)$
subgroup whose origin can be seen from the interpretation of
the $D=5$ theory as $T^5$ compactified M-theory. This $SL(6)$, 
in turn, has an $SL(3)\times SL(3)$ subgroup which can be identified with 
$SL(3)_U\times SL(3)_u$ mentioned above. Nine  charges 
represented by $p_I, q_I, r_I$, $(I=1,2,3)$ are 
within {\bf 27} of $E_6$, as can be seen by decomposing this
under $SL(6)$ and identifying them to lie within ${\bf 15}$ of 
$SL(6)$.  

The generalization of the result to $SL(5, Z)$ $U$-duality (in D=7) 
is also straightforward. One can analogously consider the case
of periodic network lattice involving 
6-string junctions (as well as other similar structures)
and define a set of five vectors, 
$\vec{\tilde{A}}_I$ $(I=1,..,5)$ similar to $\vec{a}, \vec{b}, \vec{c}$
defined earlier. Similarly one has a set of other five vectors,
$\vec{\tilde{V}}_I$, whose components are given in terms of 
quantum numbers $p_I, q_I, r_I ..$ etc., as well as 
$SL(5)/SO(5)$ moduli.
The final mass formula, now with $1/32$
supersymmetry has a form:
\begin{equation}
M_{BPS}^2 = (\vec{\tilde{V}}_I.\vec{\tilde{V}}_J) 
             (\vec{\tilde{A}}_I.\vec{\tilde{A}}_J)
          + (\vec{\tilde{V}}_I \times \vec{\tilde{V}}_J). 
            (\vec{\tilde{A}}_I \times \vec{\tilde{A}}_J).
\label{7dbpsmass}
\end{equation}
 
To conclude the discussion of the compactified non-planar 
networks as lower dimensional BPS states, we like to point out that 
several other possibilities of network compactification can be 
discussed by restricting to smaller subgroups of $U$-duality. 
For example, one can construct planar periodic 
networks of $(p,q,r)$-strings in eight dimensions, by 
considering $SL(2, Z)$ subgroups of $SL(3, Z)$.
One then has $1/4$ supersymmetric BPS states in six dimensions
after compactifying these networks on $T^2$. 
It however remains to be seen whether
one can obtain complete multiplets of the full 
duality symmetry, by combining various such possibilities 
of compactified networks.

We now discuss the application of the results to certain world-volume
theories of branes, following a similar exercise for
the case of planar IIB string networks\cite{bergman,bergman2}. 
The planar IIB configurations are of interest   
from the point of view of $1/4$ BPS dyon solutions of
$N=4$ gauge theory. These $N=4$ theories in turn are considered 
to be the linearized approximation of the world-volume theories of
D3-branes that are invariant under the $SL(2,Z)$ duality of the 
IIB theory. Moreover electric and magnetic charges 
also transform under this $SL(2, Z)$.
In eight dimensional type II theories, a similar role
is played by 2-branes which are invariant under $SL(3,Z)$ and
are known as $U2$-branes\cite{liu}. 
From the point of view of branes, this $SL(3)$ acts on charges 
originating from three different components of $N=1, D=10$ gauge
fields defining the world-volume theory.
As an example, such charges can be identified in a 2-brane of this 
type by compactifying D4-branes on $T^2$. The $SL(3, Z)$ then  
acts on three charges, originating from  the two internal components of the 
D4-brane gauge fields $(A_3, A_4)$
and a third one obtained by a Hodge-dualization
of the three-dimensional gauge fields $A_{\mu}$\cite{bhatt2}. 
In a theory of parallel multi-branes, 
these fields are expected to form appropriate adjoint representations of the 
enhanced symmetries. Existence of the $SL(3, Z)$ symmetry on 
the world-volume can also be 
argued from the point of view of heterotic strings in $D=3$. The
full duality symmetry of heterotic strings in $D=3$ is known to 
be $O(8, 24, Z)$\cite{sen3d}.
The above $SL(3, Z)$, in this picture, then 
belongs to the $SL(8, Z)$ subgroup of $O(8, 24, Z)$ , 
which transforms various components of  
ten-dimensional gauge fields, once again after Hodge-dualizations, 
in vector representations.

Then, to generalize the results of \cite{bergman}
we consider a configuration of four such branes and above
configuration of 4-string junction is formed by strings ending on 
these $U2$-branes. For example, in figure-1(a), points $(A, B, C, D)$
can be identified with the positions of these branes. 
The world-volume
of the branes is orthogonal to the three dimensional
space of strings and junctions.  
Vectors $\vec{a}, \vec{b}, \vec{c}$ described above
parameterize the vacuum expectation values of the adjoint Higgs
fields in the resulting $N=8$ supersymmetric theory. Now, to give a 
mass formula for such states while making connection with the 
work of \cite{bergman}, we choose special values for 
$p_I, q_I, r_I$ to be: $(p_1, q_1, r_1 ) = (1, 0, 0)$, 
$(p_2, q_2, r_2 ) = (0, 1, 0)$, 
$(p_3, q_3, r_3 ) = (0, 0, 1)$. In this case, BPS
mass formula (\ref{finalmass}) reduces to 
\begin{eqnarray}
M^2 = e^{-(\phi + \alpha)} |\vec{a}|^2 + 
e^{\alpha} |\vec{b}|^2 + e^{\phi} |\vec{c}|^2 + 
2 [ (\vec{a}\times \vec{b}).\hat{e}_3 e^{-\phi/2} + \cr
 (\vec{a}\times \vec{c}).\hat{e}_2 e^{-\alpha/2} +
 (\vec{b}\times \vec{c}).\hat{e}_1 e^{(\phi+\alpha)/2} ].
\label{bpsmass3d}
\end{eqnarray}
Following \cite{bergman}, we now interpret this as the mass of 
a $1/8$ supersymmetric bound state
in the world-volume theory described above. 
For this we define charges, similar to those in \cite{bergman}:
\begin{equation}
\vec{Q}_1 = e^{-(\phi + \alpha)/2}\vec{a},\>\>
\vec{Q}_2 = e^{\alpha/2}\vec{b},\>\>
\vec{Q}_3 = e^{\phi/2}\vec{c}.\>\>
\label{defcharge}
\end{equation} 
Since the role of the couplings on the world-volume theory
is played by the space-time moduli $\phi, \alpha$ etc.,  
and $(\vec{a}, \vec{b}, \vec{c})$ define the 
vacuum expectation values of scalars in this world-volume theory,
$\vec{Q}_i$'s above can be identified with 
the physical charges in this theory.
The subscripts in expression of charges now have their origins in 
fields $A_4, A_5, A_{\mu}$ mentioned above, 
whereas vector sign above them can be interpreted to be along
certain $R$-symmetry directions. 
Similar form of energy expressions, for $1/2$ BPS states
involving these fields, were observed in \cite{bhatt2}. 
However it is of interest to verify these results directly from 
the point of view of world-volume gauge theories,
following  a similar exercise in $D=4$ in \cite{fraser} and 
to further examine the properties of such bound states.

{\bf Acknowledgement: } I would like to thank Aalok 
Misra for useful communications.

\vfil
\eject


\end{document}